\documentclass[12pt]{article}

\usepackage[dvips]{graphicx}

\def\dspace{\baselineskip = 0.30in}

\def\lapproxeq{\lower .7ex\hbox{$\;\stackrel{\textstyle
<}{\sim}\;$}}
\def\gapproxeq{\lower .7ex\hbox{$\;\stackrel{\textstyle
>}{\sim}\;$}}

\begin{document}

\dspace

\begin{titlepage}
\begin{flushright}
%\preprint{
BA-05-105\\
KIAS-P05059
%March 2002\\
%}
\end{flushright}
\vskip 2cm
\begin{center}
%\title
{\Large\bf
Flat-Directions in Grand Unification \\
With $U(1)_R$ Symmetry
} \vskip 1cm {\normalsize\bf $^{(a)}$S. M.
Barr\footnote{smbarr@bxclu.bartol.udel.edu}, $^{(b)}$Bumseok
Kyae\footnote{bkyae@kias.re.kr}, and $^{(a)}$Qaisar Shafi
\footnote{shafi@bartol.udel.edu} }
\vskip 0.5cm {\it
$^{(a)}$Bartol Research Institute, Department of Physics, \\
University of Delaware, Newark, DE~~19716,~~USA
\\
$^{(b)}$School of Physics, Korea Institute for Advanced Study,
\\207-43, Cheongnyangni-Dong, Dongdaemun-Gu, Seoul 130-722,
Korea\\[0.1truecm]}
%
%%\maketitle

\end{center}
\vskip 2cm

%\date{\today}
%\pacs{PACS: 11.25.Mj, 12.10.Dm, 98.80.Cq}

\begin{abstract}

It is shown that in $SO(10)$ and $SU(5)$ models having a
$U(1)_R$ symmetry, the requirement of breaking the unified group to
the Standard Model leads to flat directions in the scalar
potential. These can lead to a ``cosmological modulus problem".
This is relevant to grand unified models of inflation, where
$U(1)_R$ symmetries are often used to insure the flatness of
the inflaton potential. A way that the modulus problem might be
avoided is discussed.

\end{abstract}

\end{titlepage}

\newpage

\section{Introduction}

Grand unified theories (GUTs) are among the most promising
extensions of the standard model. They can explain a number of
low energy phenomena, such as electric charge quantization,
neutrino oscillation, and certain fermion mass relations~\cite{abb}
as well as giving unification of gauge interactions and of matter
multiplets.

This encourages the search for the answers to various questions
in cosmology also within the GUT framework --- questions such as
the origin of cosmological inflation and baryon asymmetry.

In Refs.~\cite{hybrid}, one possible approach to these questions
was proposed in the context of supersymmetric (SUSY) GUT theory.
In particular, it was noted there that in a certain class of SUSY
models inflation is intimately associated with the spontaneous
breaking of a gauge symmetry at the GUT scale, in such a way that
$\delta T/T$ is proportional to $(M/M_{\rm Planck})^2$, where $M$
denotes the symmetry breaking scale and $M_{\rm Planck}$ ($\equiv
1.2\times 10^{19}$ GeV) denotes the Planck mass. Thus, from
measurements of $\delta T/T$, $M$ is estimated to be of order
$10^{16}$ GeV~\cite{hybrid,ns}, which is very close to the SUSY
GUT scale.

The scalar spectral index $n_s$ in these models is very close to
unity in excellent agreement with recent fits to the
data~\cite{wmap}. The vacuum energy density during inflation is of
order $10^{14}$ GeV, so that the gravitational contribution to the
quadrupole anisotropy is essentially negligible. Furthermore, the
inflaton field in this scenario eventually decays into
right-handed neutrinos, whose out of equilibrium decays lead to
the observed baryon asymmetry via (non-thermal)
leptogenesis~\cite{ls}.  It is, therefore, worthwhile to realize
this inflationary scenario within a grand unified framework.
Realistic inflationary models along this line were presented,
based on the $SU(3)_c\times SU(2)_L\times SU(2)_R\times
U(1)_{B-L}$~\cite{LR}, $SU(4)_c\times SU(2)_L\times
SU(2)_R$~\cite{422}, $SU(5)\times U(1)$~\cite{flipinf},
$SU(5)$~\cite{su5inf}, and $SO(10)$~\cite{so10inf}.

In the construction of these inflationary models, which are of the
``F-term inflation" type, a global $U(1)_R$ symmetry plays the
essential role of guaranteeing the flatness of the inflaton
potential. Indeed, $U(1)_R$ symmetry is associated with $N=1$
SUSY, and so it can reside in any $N=1$ SUSY model. This $U(1)_R$
symmetry is important also in ``D-term'' inflationary scenario for
the same reason~\cite{dterminf}. The $U(1)_R$ symmetry, however,
makes it more difficult to build SUSY GUT models with $SU(5)$,
$SO(10)$, and other unified groups. In $SU(5)$ and $SO(10)$ etc,
at least an adjoint Higgs field needs to achieve vacuum
expectation values (VEVs) of order ${\cal O}(M_G)$ in order to
break the symmetry to the standard model group. Unfortunately,
mechanisms to give VEVs to adjoint Higgs in SUSY GUT models with
$U(1)_R$ symmetry usually leave scalar fields with VEVs of order
${\cal O}(M_G)$, whose potentials are flat in the SUSY
limit.\footnote{This difficulty is readily avoided in models based
on gauge groups such as flipped $SU(5)$ and $SU(4)\times
SU(2)\times SU(2)$ which employ Higgs fields in the tensor and
bi-fundamental representations respectively~\cite{flipinf,422}.}
Such fields, which we shall call ``flat-directions,'' potentially
correspond to moduli. Moduli, here, means relatively light scalar
fields developing superlarge VEVs, whose presence are quite
problematic in cosmology. In this paper, we shall prove that
emergence of a flat-direction is not avoidable in SUSY $SO(10)$
(and also $SU(5)$) model with $U(1)_R$ symmetry in section 2, and
then propose a resolution in section 3. In section 4, we conclude.

\section{Flat-Direction in Grand Unification}

Before we prove rigorously a theorem about the connection between
$U(1)_R$ symmetry, adjoint Higgs VEVs, and flat-direction in the
context of $SO(10)$, it may be helpful first to illustrate the
problem with a very simple example.  Let us denote the adjoint
(${\bf 45}$) Higgs in $SO(10)$ by $A_H$. A simple set of terms
which would generate a VEV for it would be $W_A = M {\rm
tr}(A_H^2) + \alpha {\rm tr}(A_H^4)/M$, where $M = {\cal O}(M_G)$.
However, if there is $U(1)_R$ symmetry, then every term in $W$
must have the same $R$ charge (which we shall henceforth take to
be 1). Obviously this is not possible for the terms in $W_A$,
unless we allow one of the coefficients to be replaced by a field
that has $R \neq 0$. For example, we may consider instead $W'_A =
X {\rm tr}(A_H^2) + \alpha {\rm tr}(A_H^4)/M$, with $R(X) =
\frac{1}{2}$ and $R(A_H) = \frac{1}{4}$. That means that some
terms must exist to fix the VEV of $X$ to be of order $M_G$. One
possibility would be $W_X = S(X \overline{X} - M^2)$. The
existence of $\overline{X}$ is necessitated by the requirement of
$U(1)_R$ invariance. We immediately see the problem: this term
fixes the VEV of the product $X \overline{X}$, but leaves unfixed
the relative magnitudes of $X$ and $\overline{X}$. That is, there
remains a flat direction.

This simple example illustrates another problem that is relevant
to our later considerations. Introducing the singlet field $X$
into $W'_A$ means that there is an $F$-term equation for $X$ to be
satisfied. In this case, the equation $F_X =0$ implies that ${\rm
tr}(A_H^2) + S \overline{X} = 0$. Since, by hypothesis $A_H \neq
0$, and by the $F_{\overline{X}} = 0$ condition $S = 0$, one has a
contradiction. A simple way to avoid this difficulty would be to
have two adjoints, one of which has vanishing VEV. For example,
one might consider $W^{\prime \prime}_A = {\rm tr}(A \; A_H) +
\alpha {\rm tr}(A \; A_H^3)/ M$, where $\langle A \rangle = 0$.
The $U(1)_R$ invariance is ensured by the choices $R(A) = 1$ and
$R(A_H) = 0$. The equation $F_{A_H} = 0$ is satisfied by the
vanishing of $A$, while $F_A$ fixes $A_H$. This seems to obviate
the difficulty encountered before, in that we have not had to
introduce the singlet $X$ to satisfy $U(1)_R$ invariance. However,
the two terms in $W^{\prime \prime}_A$ are insuffiencient to give
mass to all the fields in $A$ and $A_H$. In particular, there are
several color-triplet fields that remain light, and that would be
disastrous for the running of the gauge couplings. All fields in
the adjoints can be made heavy if an additional term is
introduced: $X {\rm tr}(A^2) + M {\rm tr}(A \; A_H) + \alpha {\rm
tr}(A \; A_H^3)/M$. Note the crucial point that the singlet field
$X$ has had to be introduced again to satisfy $U(1)_R$ invariance.

The lessons of these simple examples can in fact be generalized to
a theorem, which we now state, and shall then prove: {\it In
$SO(10)$, if the superpotential has $U(1)_R$ symmetry, and if the
adjoint Higgs fields have ${\cal O}(M_G)$ VEVs and contain no
goldstone or pseudo-goldstone components, then there must exist
fields with $O(M_G)$ VEVs that have a flat direction.} In
short, fixing the VEVs of the adjoints when there is $U(1)_R$
symmetry leads to ``flat-directions".

%\noindent
%{\bf The Proof:}

We shall prove the theorem first in the simplified context of
$SO(10)$ models whose Higgs sector consists of the adjoints $A_H$
and $A$ and some number of singlets. We shall distinguish between
two kinds of fields that appear in the Higgs superpotential: those
that have ${\cal O}(M_G)$ VEVs we shall call ``Higgs fields",
while those that have vanishing VEVs in the supersymmetric limit
we shall call ``null fields". The Higgs fields will therefore
consist of $A_H$ and some number $N$ of singlets that we shall
denote $\phi_i$, $i=1, ..., N$. The null fields will consist of
$A$ (possibly) and some number $N_S$ of singlets that we shall
denote $S_a$, $a = 1, ..., N_S$.)

We shall try to fix the VEVs of all these fields (i.e. try to
avoid the existence of flat directions), but shall find that this
cannot be done if the superpotential has $U(1)_R$ symmetry. One
can distinguish three kinds of terms in the Higgs superpotential:
Type 0 terms contain no null fields; Type 1 terms are first order
in null fields; and Type 2 are second order or higher in null
fields. The first thing to be noted is that terms of Type 2 in the
superpotential make no contribution to $F$ terms at the minimum
and thus have no effect on the minimization of the Higgs potential
or the VEVs of the fields. Therefore, in fixing the VEVs we need
only consider terms of Type 0 and Type 1. First we shall deal with
the case where $W$ contains no Type 0 terms.

\subsection{Case with no Type 0 terms}

If we assume that the superpotential contains no Type 0 terms, it
can be taken to have the form:

\begin{equation}
W = W_S + W_A + W_2,
\end{equation}

\noindent
where $W_2$ represents Type 2 terms that can be neglected,

\begin{equation}
W_S = \sum_{a=1}^{N_S} S_a  P_a(\phi_i, A_H),
\end{equation}

\noindent
and

\begin{equation}
W_A = X_1 {\rm tr}(A^2) + X_2 {\rm tr}(A \; A_H) + X_3 {\rm tr}(A
\; A_H^3) + h.o.
\end{equation}

\noindent The $P_a$ in Eq. (2) are polynomials in the Higgs
fields. The factors $X_1$, $X_2$ and $X_3$ in Eq. (3) represent
either singlet Higgs fields ($\phi_i$) or products of Higgs fields
($\phi_i$ and $A_H$) divided by appropriate powers of $M_G$ to
make the terms dimensionally correct. We include the term $X_1
{\rm tr}(A^2)$ even though it is of Type 2 and does not affect the
minimization, because it must be there to give mass to all the
color-triplet fields in the adjoints (as noted earlier), and
because it shall play a crucial role in the proof by constraining
the $U(1)_R$ charges of fields. The ``$h.o.$" in Eq. (3)
represents terms containing ${\rm tr}(A \; A_H^n)$ with $n$ larger
than 3. These can be included but would not affect the analysis.

Each of the three terms in Eq. (3) must have $R = 1$ and must be
neutral under all local $U(1)$ symmetries. Consider, then, the
product

\begin{equation}
\Pi \equiv [X_1]^{-1} [X_2]^3 [X_3]^{-1}.
\end{equation}

\noindent This product obviously has the same charges as the
following product of terms in Eq. (3): $[X_1 {\rm tr}(A^2)]^{-1}
[X_2 {\rm tr}(A \; A_H)]^3 [X_3 {\rm tr}(A \; A_H^3)]^{-1}$. Since
each term in Eq. (3) must have gauge charge zero and $R=1$, it
must be that $\Pi$ has $R = -1 + 3 - 1 = 1$ and is neutral under
all local $U(1)$ symmetries. Moreover, since it is made up of
powers of the Higgs fields $\phi_i$ and $A_H$, it has a VEV that
is ${\cal O}(M_G^3)$. The crucial question will be whether such a
product can exist if there are no flat directions.

There are only two kinds of terms in the Higgs potential available to fix
the VEVs of the $\phi_i$ and $A_H$ fields: $D$-terms corresponding to whatever
local $U(1)$ symmetries exist in the model, and $F$-terms corresponding to the
null fields $S_a$ and $A$. (It is clear from the fact that
there are no Type 0 terms in the Higgs superpotential that
$F_{\phi_i}$ and $F_{A_H}$ automatically vanish, since they
necessarily have at least one power of a null field. That means that only
the $F$-terms $F_{S_a}$ and $F_A$ affect the minimization.)

Let the gauged $U(1)$ groups be denoted $U(1)_K$, $K =
1,...,N_D$. The charges of the Higgs fields $\phi_i$, $i=1,...,N$, and
$A_H$ under the group $U(1)_K$ can be thought of as a vector in an
$(N+1)$-dimensional space: $\vec{Q}^K = (Q^K(\phi_1), ...,Q^K(\phi_N),
Q^K(A_H))$. There are $N_D$ such vectors, one for each $U(1)_K$. However, it
may be that not all of these vectors are independent. Suppose that
$\tilde{N}_D$ of them are independent. Then the $D$-terms provide
$\tilde{N}_D$ independent conditions on the VEVs of the Higgs fields.
We can also think of the $R$ charges of the fields as forming an
$(N+1)$-component vector in the same space: $\vec{R} = (R(\phi_1), ...,
R(\phi_N), R(A_H))$.

There are $(N_S + 1)$ conditions (of the form $P_a = 0$) on the VEVs of
$\phi_i$ and $A_H$ coming from $F_{S_a} =0$, $a=1,...,N_S$, and $F_A =0$.
It is easily seen that these $F$-term conditions must be independent.
(If some $P_a$ were expressible as linear combinations of others, that
would obviously mean that there were exact relationships among
some of the coefficients in the $P_a$, which would require fine-tuning.)
Thus, there are
a total of $(\tilde{N}_D + N_S + 1)$  conditions on the VEVs of $(N+1)$
fields. Consequently, if all Higgs field
VEVs are to be fixed in the supersymmetric
limit, i.e. if there are to be no flat directions, it must
be that $\tilde{N}_D + N_S \geq N$.
We will now show that this
would imply that the $U(1)$ charges of Higgs fields are so highly constrained
that no product $\Pi$ of them can be constructed that has $R(\Pi) = 1$ and
$Q^K(\Pi) = 0$, as needed to write down the terms in Eq. (3).

Consider one of the polynomials $P_a$ in Eq. (2). In order to satisfy
the relation $-F^*_{S_a} = P_a = 0$, the polynomial $P_a$ must have
$p$ terms, where $p \geq 2$, since each term in $P_a$ is a product
of fields with non-zero VEVs. Then the requirement that these $p$
terms all have the same charge under a $U(1)$ gives $p-1$ homogeneous
linear relations on the charges of the Higgs fields under that $U(1)$.
(Note that this is true also for $U(1)_R$.) Consequently,
{\it for any $U(1)$ group}, the terms in $W_S$ give {\it at least}
$N_S$ homogeneous linear relations on the charges of the Higgs
fields under that group.

Moreover, for any $U(1)$, the requirement that the three terms displayed
in Eq. (3) have the same charge under that group (in particular $0$ for a
gauge $U(1)$ and $1$ for $U(1)_R$) gives two homogeneous
linear relations that must be satisfied by the charges of the fields $A$,
$A_H$, and $\phi_i$ under that $U(1)$. Eliminating the charge of the
null field $A$ from these relations, one is left with exactly one homogeneous
linear relation on the charges of the the Higgs fields under the $U(1)$.
The ``other terms" in Eq. (3) may give additional linear relations.

We thus have that the components of any ``charge vector" (whether it
be $\vec{Q}^K$ or $\vec{R}$) must satisfy at least $N_S +1$ homogeneous
linear relations. Therefore, all the charge vectors lie in a subspace
that has dimension $\leq (N+1) - (N_S +1) = (N - N_S)$. Since it has
already been shown that $\tilde{N}_D + N_S \geq N$ (in order to fix all
the VEVs), the subspace in which all the
charge vectors lie must have dimension $\leq \tilde{N}_D$.
However, the number of vectors $\vec{Q}^K$ is $\tilde{N}_D$ and they are
all independent, so they must span this subspace. Consequently, $\vec{R}$,
which also lies in this subspace, must be expressable
as a linear combination of the $\vec{Q}^K$: i.e. $\vec{R} =
\sum_K \alpha_K \vec{Q}^K$. That means that any product of powers of the
Higgs fields $\phi_i$ and $A_H$ that has vanishing charge under all the
gauge groups $U(1)_K$ must also have vanishing $R$ charge. However, this
contradicts the requirement for writing down the terms in Eq. (3), namely
that some product $\Pi$, given in Eq. (4), has $Q^K(\Pi) = 0$ and
$R(\Pi) = 1$. We have thus proven the theorem in the case where $W$ contains no
Type 0 terms. Now we turn to the case where there are Type 0 terms.

\subsection{The case with Type 0 terms}

Before doing the general case, consider a few simple examples. Let
$W_0$ comprise the Type 0 terms in $W$, and let them all depend on
only one field, a gauge-singlet Higgs field that we will denote $H$. If
$W_0$ contains only a single monomial in $H$, say $H^p$, then the
condition $F_H = 0$ will force $\langle H \rangle$ to vanish,
which is a contradiction since $H$ would then be a null field
appearing in $W_0$, which by definition contains no such fields.
There must therefore be at least two different monomials in $W_0$,
e.g. $W_0 = a H^p + b H^q$, with $p \neq q$, and $a$ and $b$ being
some coefficients. However, it is then obviously impossible to
make both terms in $W_0$ have $R = 1$. This illustrates the
general difficulty that if there are enough terms in $W_0$ to fix
the VEVs of all the Higgs that it contains, then there are too
many constraints on the $R$ charges to be satisfied. To put it the
other way, if the $R$ charges can be consistently assigned, there
must be at least one VEV that does not get fixed, i.e. a flat
direction. A simple illustration of this is the following. Let
$W_0 = H_1 H_2 H_3 + M H_2 H_3 + M^2 H_3 + (1/M) H_1 H_2^2 H_3$.
Then $R$ invariance is satisfied by assigning $R(H_1) = R(H_2) =0$
and $R(H_3) =1$. Moreover, $F_{H_1} = 0$ gives $H_2 = -M$; and
$F_{H_2} = 0$ gives $H_1 = M$. However, there is a flat direction,
since the remaining condition $F_{H_3} =0$ gives $H_1 H_2 + M H_2
+ M^2 + (1/M) H_1 H_2^2 = 0$, which is automatically satisfied
independently of the value of $H_3$. Note that the flat direction
corresponds to the $R$ charge assignments. (That is, only $H_3$ has
a non-zero $R$ charge, and the flat direction is the $H_3$ direction
in field space.) This result generalizes, as we now show.

Let $W_0$ depend on $N_0$ gauge-singlet Higgs fields, which we will denote
$H_{\alpha}$, $\alpha = 1,...,N_0$. Let $W_0$ have the form

\begin{equation}
W_0 = \sum_{n=1}^{N_T} T_n (H_{\alpha}), \;\; n = 1,...,N_T.
\end{equation}

\noindent
where each term $T_n$ is a monomial in the Higgs fields:
$T_n = c_n (H_1)^{a_{1n}}...(H_{N_0})^{a_{N_0 n}}$. There are $N_0$
conditions $F_{H_{\alpha}} = 0$. It is convenient to write them in the form
(no sum over $\alpha$):

\begin{equation}
H_{\alpha} \frac{\partial}{\partial H_{\alpha}} W_0 = 0, \;\;
\alpha = 1,...,N_0.
\end{equation}

\noindent
Obviously the other terms in $W$ besides $W_0$ do not contribute to these
conditions since they all contain at least one null field. The operator
$H_{\alpha} \frac{\partial}{\partial H_{\alpha}}$ acting on any term
$T_n$ in Eq. (5) just multiplies that term by a number, $a_{\alpha n}$.
Consequently, the equations given in Eq. (6) are just homogeneous linear
equations in the terms $T_n$:
\begin{equation}
H_{\alpha} \frac{\partial}{\partial H_{\alpha}} W_0 = \sum_n
H_{\alpha} \frac{\partial}{\partial H_{\alpha}} T_n = \sum_n
a_{\alpha n} T_n = 0.
\end{equation}

\noindent
We may assume that these $N_0$ equations are all linearly independent.
For, if they were not, it would mean that
some of the fields $H_{\alpha}$ only appeared in $W_0$ in certain product
combinations. (For example, if $H_1$ and $H_2$ only appeared in the
combination $H_1^p H_2^q$, then the $F_{H_1}$ equation and $F_{H_2}$
equation would be proportional.) We could then take those product combinations
to be new singlet fields $H'_{\alpha}$. The equations that
resulted from differentiating $W_0$ with respect to these new fields would be
linearly independent. We can assume that we have already performed such a
reduction to get Eq. (5), so that the $N_0$ equations in Eq. (6) are
linearly independent.

Now, suppose that that all the terms in $W_0$ have $R=1$ and that $W_0$
contains enough terms that there is a solution to Eq. (6) with all the
$T_n \neq 0$. (We have already given an example of this with $N_0 = 3$ above.)
Let the fields have values $\overline{H}_{\alpha}$ at this solution. Consider
scaling this solution by powers of some complex number $\lambda$ in the
following way: $H_{\alpha} =
\lambda^{R(H_{\alpha})} \overline{H}_{\alpha}$. Since each term in $W_0$ has
$R(T_n) = 1$, every term $T_n$ will scale by a factor of exactly $\lambda$.
Therefore, the equations given in Eq. (7) are all still satisfied,
which in turn implies that the direction parametrized by $\lambda$ is $F$-flat.
(And this general result is verified in the example given above, where the
scaling would affect only the field $H_3$, which is indeed the flat direction.)

We have assumed that the fields $H_{\alpha}$ are singlets.
However, the same argument easily generalizes to non-singlet
fields. For example, suppose we allow one of these fields to be an
adjoint $A_H$. Only traces of an even power of $A_H$ can appear
because the adjoint of $SO(10)$ is antisymmetric. We may write all
the traces of powers of $A_H$ that appear in $W_0$ as combinations
of ${\rm tr}(A_H^2)$ and ratios of the form $R_p \equiv {\rm
tr}(A_H^{2p})/({\rm tr}(A_H^2))^p$. Now given a certain form of
the VEV of $A_H$ these ratios $R_p$ just give numbers. Moreover,
in the equation $A_H \frac{\partial W_0}{\partial A_H} = 0$ from
Eq. (6) the factors of $R_p$ make no difference, since $A_H
\frac{\partial R_p}{\partial A_H} =0$.) Thus, we may effectively
take the $R_p$ to be pure numbers and ${\rm tr}(A_H^2)$ to be a
singlet field in the argument we made before.

Now, all we have shown up to this point is that there is a direction that
is undetermined (i.e. flat) by the conditions $F_{H_{\alpha}} =0$. That does
not imply that when all the $F$ and $D$ terms are taken into account
there must be a flat direction. To prove the theorem requires a few more
steps. First, let us prove that there are at least $N_0$ independent
terms in $W_0$. (A set of monomials in the fields $H_{\alpha}$ is independent
if none of them can be expressed as a product of powers of the others.)
Let us suppose that of the $N_T$ terms in $W_0$, $M$ are
independent. Then the remaining $(N_T -M)$ terms can be written in terms
of the independent ones as follows:
\begin{equation}
W_0 = \sum_{n=1}^M T_n + \sum_{\ell = M+1}^{N_T} c_{\ell} \left(
T_1^{p_{1 \ell}} ... T_M^{p_{M \ell}} \right).
\end{equation}

\noindent
Then Eq. (6) becomes
\begin{equation}
0 = \sum_{n=1}^M a_{\alpha n} T_n + \sum_{\ell = M+1}^{N_T}
\sum_{n=1}^M a_{\alpha n} p_{n \ell} T_{\ell} = \sum_{n=1}^M
a_{\alpha n} \left( T_n + \sum_{\ell = M+1}^{N_T} p_{n \ell}
T_{\ell} \right).
\end{equation}

\noindent
This is a set of $N_0$ independent linear equations in the $M$
quantities $\tilde{T}_n \equiv T_n + \Sigma_{\ell = M+1}^{N_T}
p_{n \ell} T_{\ell}$. There can be no solution unless $M \geq N_0$, which
is what we wanted to show. We are now in a position to complete the proof
of the theorem.

Consider the case where the Higgs superpotential consists of
the terms given
in Eqs. (1), (2), (3), and (5), where the $N_0$ fields $H_{\alpha}$ appearing
in Eq. (5) are a subset of the $(N +1)$ fields $\phi_i$ and $A_H$.
As was shown previously, there are $(N_S +1)$ conditions on the VEVs
coming from $F_{S_a} =0$ and $F_A = 0$, and $\tilde{N}_D$ conditions
coming from the $D$ terms. And we have shown that there are no more than
$(N_0 -1)$ conditions on the VEVs coming from $F_{H_{\alpha}} =0$ (not $N_0$
because of the direction that is left flat by those conditions).
Altogether, then, there are no more than $(N_S + N_0 + \tilde{N}_D)$
conditions, and these must be sufficient to fix the VEVs of the $(N+1)$ fields
$\phi_i$ and $A_H$. That implies that $(N_S + N_0 + \tilde{N}_D) \geq (N+1)$.

On the other hand, we showed that coming from the terms in $W_S$ and $W_A$
there are $(N_S + 1)$ homogeneous linear relations on the charges of the Higgs
fields under any $U(1)$ group. In addition, because there are at least
$N_0$ independent terms in $W_0$, the fact that these terms must all have
the same charges (i.e. 0 for gauge $U(1)$ and 1 for $U(1)_R$), yields
at least another $(N_0 -1)$ homogeneous linear relations on the
charges of the Higgs fields under any $U(1)$. Altogether, then, there are
at least $(N_S + N_0)$
homogeneous linear relations on the charges of the $(N+1)$ Higgs fields.
That means that each ``charge vector" lies in a subspace of dimension
$\leq (N+1 - N_S - N_0) \leq \tilde{N}_D$. That implies that the
$\tilde{N}_D$ independent
vectors $\vec{Q}^K$ must span this subspace, and that therefore
$\vec{R}$ is a linear combination of the $\vec{Q}^K$. As before, that
means that the product $\Pi$ in Eq. (4) cannot have both $Q^K(\Pi) = 0$
and $R(\Pi) =1$ as required to write the terms in Eq. (3).

There is one final case to consider. Suppose that the Higgs superpotential
consists of $W_S$ and $W_0$, but has no terms of the form $W_A$ given in
Eq. (3). (So there is no $A$ field.) That is, suppose the VEV of $A_H$ is
set by $W_0$, $W_S$ and $D$ terms, rather than by $W_A$.
In that case, there are $N_S + (N_0 -1)
+ \tilde{N}_D$ conditions on the $(N+1)$ Higgs VEVs, so that $(N_S + N_0
+ \tilde{N}_D) \geq (N+2)$. On the other hand,
the terms in $W_S$ and $W_0$ yield at least $N_S + (N_0 -1)$ independent
linear relations on the charges of the Higgs under any $U(1)$ group.
Thus the charge vectors lie in a subspace of dimension
$(N+1) - (N_S + N_0 -1) \leq \tilde{N}_D$.
The argument then proceeds as before.

The foregoing argument applies to $SU(5)$ as well as $SO(10)$. In
$SU(5)$, one is allowed to have a cubic coupling of an adjoint
Higgs, so that the breaking of $SU(5)$ can be achieved with the
terms $M {\rm tr}(A A_H) + {\rm tr}(A A_H^2)$, where $A$ and $A_H$
are in the ${\bf 24}$. However, it is easy to show that these
terms are not enough to give mass to all the un-eaten modes of the
adjoints. Thus, to avoid goldstone modes, one must add a term like
${\rm tr}(A^2)$, as in the $SO(10)$ case. The reasoning then
proceeds in complete analogy with the $SO(10)$ case, the only (and
insignificant) exception being that there is a cubic rather than a
quartic term in $W_A$.

\section{Flat-Direction and Modulus}

So far we have proven that in SUSY GUTs based on $SO(10)$ and $SU(5)$
(and other groups that require an adjoint to get a VEV to break to
the Standard Model) $U(1)_R$ symmetry leads to flat directions.
A direction that is flat in the supersymmetric limit can lead to a
``cosmological modulus problem". The point is that soft SUSY-breaking terms
will give a tiny mass in the flat direction and pick out some point in
that direction as the true minimum. However, in the early universe this
``modulus" field may find itself ``initially" far from the true minimum ---
indeed, one might typically expect that it finds itself a distance of order
$M_G$ from the minimum. In that case, after SUSY breaking, the modulus field
would oscillate about its true minimum, and the energy in these
oscillations would have disastrous consequences for standard
Big Bang cosmology.

However, even if there is a flat direction, a cosmological catastrophe
would be avoided if the ``modulus" field happened ``initially" to
find itself at the true minimum. What we mean by the ``initial" value of
the modulus field is its value after inflation. This value is determined
by temperature-dependent terms in the effective
potential that lift the degeneracy in the flat direction. On the other hand,
the minimum at low temperature is determined by the soft SUSY-breaking terms.
The question is whether the high-temperature terms select out the
same point along the flat direction as the low-energy SUSY-breaking
terms. If so, there is no problem.

Let us illustrate with a toy model how the modulus might find
itself at the true minimum initially. Consider a superpotential of
three scalar superfields $S$, $X$ and $Y$ of the form $W =
S(\frac{1}{M_1^{(a+b-2)}} X^a Y^b - M_2^2)$. The flat direction is
given by $Y = (M_2^2 M_1^{(a+b-2)})^{1/b} X^{-a/b}$. Suppose the
soft SUSY-breaking terms are ``universal", i.e. of the form $m_0^2
(|X|^2 + |Y|^2 + ...)$. (Ignore $A$ terms.) This would pick out
the minimum $X = (a/b)^{b/2(a+b)} (M_2^2 M_1^{(a+b-2)})^{1/(a+b)}$
and $Y = \sqrt{b/a} X$. Now, consider the situation during
inflation. If the K\"{a}hler potential is of the minimal form,
then during inflation all scalars can have an equal contribution
to their mass-squared given by $M_{\rm infl}^2 (|X|^2 + |Y|^2 +
...)$, with $M_{\rm infl} \sim \Lambda^2/M_P$. $\Lambda$ is the
vacuum energy density during inflation and $M_P$ ($= 2.4 \times
10^{18}$ GeV) is the reduced Planck mass. If $M_{\rm infl}$ is
large enough, the fields $X$ and $Y$ will be trapped at the origin
during inflation. However, the crucial point is that both the
$M_{\rm infl}$ term and the soft SUSY-breaking terms that dominate
at low energy have the same ``universal form". That means that as
$X$ and $Y$ evolve away from zero after inflation they will
maintain the ratio $Y = \sqrt{b/a} X$ and go directly to the true
minimum. In other words, because the high-$T$ and low-$T$
mass-squared terms have the same form, the modulus finds itself
``initially" at the true minimum, thus avoiding the cosmological
difficulties coming from energy trapped in oscillations about that
minimum.

Somewhat more generally, assume the flat-direction
is one dimensional line in two dimensional field space $(\psi_1,
\psi_2)$, as shown in Fig. 1.
\vskip 0.5cm
\begin{center}
\includegraphics[width=120mm]{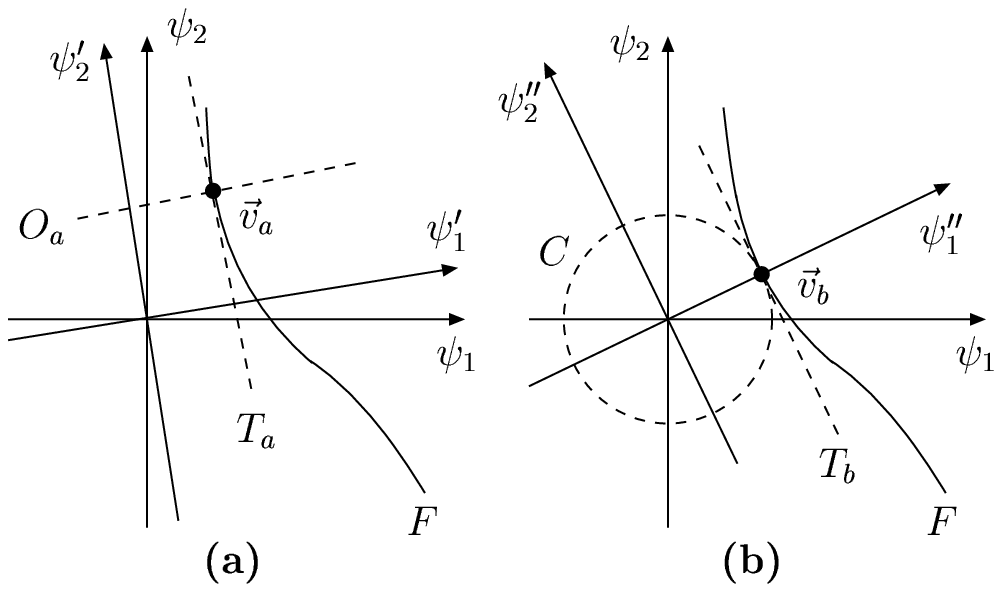}
\end{center}
FIG. 1: By including SUSY breaking soft terms in the scalar
potential, VEVs of a flat-direction $F$ are assumed to be
determined at $\vec{v}_a$ (a) and $\vec{v}_b$ (b), respectively.
We neglect deformations of $F$ by soft terms.
\vskip 0.5cm

\noindent The line ``F'' in Fig. 1(a) and 1(b) denotes the
flat-direction, along which the scalar potential vanishes in the
SUSY limit. By including SUSY breaking soft terms, the vacuum
state is assumed to be fixed on $\vec{v}_a$ [in (a)] or
$\vec{v}_b$ [in (b)]. Soft terms would deform the flat-direction
just by ${\cal O}(m_{3/2})$, which we may neglect.
Once the vacuum is determined, the physical mass eigenstates on that
vacuum state can be found. $\psi_{1,2}'$ and
$\psi_{1,2}''$ correspond to the mass eigenstates in Fig.
1(a) and 1(b), respectively. Since $\psi_2'$ and $\psi_2''$ are
parallel to the tangent lines $T_a$ and $T_b$ on $\vec{v}_a$ and
$\vec{v}_b$, their masses are ${\cal O }(m_{3/2})$ [zero
in the SUSY limit].  On the other hand, the masses of $\psi_1'$ and
$\psi_1''$, which are orthogonal to $\psi_2'$ and $\psi_2''$
respectively, should be ${\cal O}(M_G)$.
The important difference between (a) and (b) arises from the VEVs of
$(\psi_1',\psi_2')$ and $(\psi_1'',\psi_2'')$. In case (a), the VEVs
of both $\psi_1'$ and $\psi_2'$ are ${\cal O}(M_G)$, so that
the oscillations of the light scalar $\psi_2$ would generally give rise
to the cosmological modulus problem.
However, in type (b), the VEV of the light field $\psi_2''$
vanishes, as can be seen from Fig. 1(b), and no modulus problem results, as
long as the ``initial" value of $\psi_2''$ also vanishes.
Note that the
tangent line ``$T_b$" of the flat-direction ``$F$'' at
$\vec{v}_b$ is also a tangent line to the circle ``$C$'' whose
center is located at the origin. If the terms that lift the degeneracy
of the flat direction $F$ are of ``universal" form, then their effect, of
course, is to minimize the distance to the origin,
If both the high-temperature and the low-temperature
terms that lift the degeneracy of $F$ are minimal in form, then
the fields will evolve along the straight line from the origin to
the point on $F$ that is closest to the origin. As can be seen from
Fig. 1(b), this means that the ``modulus" mode is never excited, and
no cosmological modulus problem arises.

The toy model involving the superfields
$S$, $X$, and $Y$ that we gave at the beginning of this section is
a realization of case (b). (As the temperature falls in that example, the
fields evolve from the origin directly in a straight line $Y =
\sqrt{b/a} X$ toward the true
minimum, so that the field we call $\psi_2''$ in Fig. 1(b)
always vanishes.)

It remains to ask whether the high-temperature and low energy
effective mass terms
can be of minimal form. If the K${\rm\ddot{a}}$hler potential
itself has a minimal form, then this can be the case. Moreover,
in gauge-mediated SUSY breaking scenario, the ``A-terms'' (which we neglected
in the above analysis) are generally small.
And even in some special cases of gravity mediation the ``A-term''
contributions to the scalar potential $V(\psi_1, \psi_2)$
may be suppressed~\cite{so10inf}. Suppose that the VEV of
the superpotential in which $\psi_1$ and $\psi_2$ are
involved ($\equiv \langle W(\psi_1,\psi_2)\rangle$) is
cancelled by another term ($\equiv \langle W_c\rangle$). If the
dimensions of $\langle W(\phi)\rangle$ and $\langle W_c\rangle$
are the same, then the ``A-term'' contribution to
$V(\psi_1,\psi_2)$ with the minimal K${\rm\ddot{a}}$hler
potential would effectively also cancel at the leading order.

The above discussion can obviously be extended to the more general case
with any number of fields.  If the
tangent space at a vacuum point of an N-dimensional flat-direction
coincides with the tangent space at that point of the sphere whose
center is at the origin, one has a situation of type (b).

\section{Conclusion}

We have proved a theorem that in models based on $SO(10)$ and
$SU(5)$ the existence of a $U(1)_R$ symmetry together with the
requirement of breaking the unified
group to the standard model leads to the existence of flat
directions in the scalar potential in the SUSY limit.
Such light modes (which generally obtain masses of order
$m_{3/2}$ when supersymmetry breaks) can lead to the well known
``cosmological modulus problem". These observations are relevant to
the problem of building grand unified models of inflation, where a
$U(1)_R$ symmetry is typically required to ensure the flatness of
the inflaton potential. We have noted in the previous section that
there is a way that the cosmological modulus problem can in principle
be avoided, even if there are flat directions, if the modulus modes
find themselves at the true minimum of their potential after
inflation. We have discussed conditions under which this may be the case.

\vskip 0.3cm \noindent {\bf Acknowledgments}

\noindent S.M.B and Q.S. are partially supported by the DOE under
contract No. DE-FG02-91ER40626.

%%%%%%%%%%%%%%%%%%%%%%%%%%%%%%%%%%%%%%%%%%%%%%%%%%%%%%%%%%%%%%%%%%%%%%%%%%%


\begin{thebibliography}{99}

%\bibitem{seesaw} M.~Gell-Mann, P.~Ramond and R.~Slansky,
%  %``Complex Spinors And Unified Theories,''
%Print-80-0576 (CERN);
%%\href{http://www.slac.stanford.edu/spires/find/hep/www?r=print-80-0576\%2F(cern)}{SPIRES entry}
% T.~Yanagida,
%  %``Horizontal Gauge Symmetry And Masses Of Neutrinos,''
%%\href{http://www.slac.stanford.edu/spires/find/hep/www?irn=518573}{SPIRES entry}
%{\it In Proceedings of the Workshop on the Baryon Number of the
%Universe and Unified Theories, Tsukuba, Japan, 13-14 Feb 1979};
% R.~N.~Mohapatra and G.~Senjanovic,
%  %``Neutrino Mass And Spontaneous Parity Nonconservation,''
%  Phys.\ Rev.\ Lett.\  {\bf 44} (1980) 912;
%  %%CITATION = PRLTA,44,912;%%
%  S.~L.~Glashow,
%  %``The Future Of Elementary Particle Physics,''
%HUTP-79-A059
%%\href{http://www.slac.stanford.edu/spires/find/hep/www?r=hutp-79-a059}{SPIRES entry}%
%{\it Based on lectures given at Cargese Summer Inst., Cargese,
%France, Jul 9-29, 1979}

\bibitem{abb} For instance, see C.~H.~Albright and S.~M.~Barr,
%``Fermion masses in SO(10) with a single adjoint Higgs field,''
Phys.\ Rev.\ D {\bf 58} (1998) 013002 [arXiv:hep-ph/9712488];
C.~H.~Albright, K.~S.~Babu and S.~M.~Barr,
%``A minimality condition and atmospheric neutrino oscillations,''
Phys.\ Rev.\ Lett.\  {\bf 81} (1998) 1167 [arXiv:hep-ph/9802314];
C.~H.~Albright and S.~M.~Barr,
%``Predicting quark and lepton masses and mixings,''
Phys.\ Lett.\ B {\bf 452} (1999) 287 [arXiv:hep-ph/9901318];
S.~M.~Barr,
%``A Different seesaw formula for neutrino masses,''
Phys.\ Rev.\ Lett.\  {\bf 92} (2004) 101601
[arXiv:hep-ph/0309152]; S.~M.~Barr and B.~Kyae,
%``A general analysis of corrections to the standard see-saw formula in grand
%unified models,''
Phys.\ Rev.\ D {\bf 70} (2004) 075005 [arXiv:hep-ph/0407154].

\bibitem{hybrid} G.~R.~Dvali, Q.~Shafi and R.~K.~Schaefer,
%``Large Scale Structure And Supersymmetric Inflation Without Fine Tuning,''
Phys.\ Rev.\ Lett.\  {\bf 73} (1994) 1886 [arXiv:hep-ph/9406319].
For a review and additional references, see G.~Lazarides,
%``Inflationary cosmology,''
Lect.\ Notes Phys.\  {\bf 592} (2002) 351 [arXiv:hep-ph/0111328].
See also D.~H.~Lyth and A.~Riotto,
%``Particle physics models of inflation and the cosmological density
%perturbation,''
Phys.\ Rept.\  {\bf 314} (1999) 1 [arXiv:hep-ph/9807278].

\bibitem{ns}
V.~N.~Senoguz and Q.~Shafi,
%``Testing supersymmetric grand unified models of inflation,''
Phys.\ Lett.\ B {\bf 567} (2003) 79 [arXiv:hep-ph/0305089];
%
%\bibitem{dec}
V.~N.~Senoguz and Q.~Shafi,
%``Reheat temperature in supersymmetric hybrid inflation models,''
Phys.\ Rev.\ D {\bf 71} (2005) 043514 [arXiv:hep-ph/0412102].

\bibitem{wmap} D.~N.~Spergel {\it et al.},
 %``First Year Wilkinson Microwave Anisotropy Probe (WMAP) Observations:
%Determination of Cosmological Parameters,''
Astrophys.\ J.\ Suppl.\  {\bf 148} (2003) 175
[arXiv:astro-ph/0302209]; C.~L.~Bennett {\it et al.},
 %``First Year Wilkinson Microwave Anisotropy Probe (WMAP) Observations:
%Preliminary Maps and Basic Results,''
Astrophys.\ J.\ Suppl.\  {\bf 148} (2003) 1
[arXiv:astro-ph/0302207]; H.~V.~Peiris {\it et al.},
 %``First year Wilkinson Microwave Anisotropy Probe (WMAP) observations:
%Implications for inflation,''
Astrophys.\ J.\ Suppl.\  {\bf 148} (2003) 213
[arXiv:astro-ph/0302225]. See also G.~F.~Smoot {\it et al.},
%``Structure in the COBE DMR first year maps,''
Astrophys.\ J.\  {\bf 396} (1992) L1; C.~L.~Bennett {\it et al.},
 %``4-Year COBE DMR Cosmic Microwave Background Observations: Maps and Basic
%Results,''
Astrophys.\ J.\  {\bf 464} (1996) L1 [arXiv:astro-ph/9601067].




\bibitem{ls} M.~Fukugita and T.~Yanagida,
%``Baryogenesis Without Grand Unification,''
Phys.\ Lett.\ B {\bf 174} (1986) 45.  For non-thermal
leptogenesis, G.~Lazarides and Q.~Shafi,
%``Origin Of Matter In The Inflationary Cosmology,''
Phys.\ Lett.\ B {\bf 258} (1991) 305.

\bibitem{LR} G.~R.~Dvali, G.~Lazarides and Q.~Shafi,
 %``mu problem and hybrid inflation in supersymmetric SU(2)L x SU(2)R x
%U(1)B-L,''
Phys.\ Lett.\ B {\bf 424} (1998) 259 [arXiv:hep-ph/9710314].

\bibitem{422} R.~Jeannerot, S.~Khalil, G.~Lazarides and Q.~Shafi,
%``Inflation and monopoles in supersymmetric SU(4)c x SU(2)L x SU(2)R,''
JHEP {\bf 0010} (2000) 012 [arXiv:hep-ph/0002151]; S.~F.~King and
Q.~Shafi,
%``Minimal supersymmetric SU(4) x SU(2)L x SU(2)R,''
Phys.\ Lett.\ B {\bf 422} (1998) 135 [arXiv:hep-ph/9711288].

\bibitem{flipinf}   B.~Kyae and Q.~Shafi,
  %``Flipped SU(5) predicts delta(T)/T,''
  arXiv:hep-ph/0510105.

\bibitem{su5inf}
%
%L.~Covi, G.~Mangano, A.~Masiero and G.~Miele,
%%``Hybrid inflation from supersymmetric SU(5),''
%Phys.\ Lett.\ B {\bf 424} (1998) 253 [arXiv:hep-ph/9707405];
%T.~Watari and T.~Yanagida,
%%``GUT phase transition and hybrid inflation,''
%Phys.\ Lett.\ B {\bf 589} (2004) 71 [arXiv:hep-ph/0402125];
%%
  B.~Kyae and Q.~Shafi,
  %``delta(T)/T and neutrino masses in SU(5),''
  Phys.\ Lett.\ B {\bf 597} (2004) 321
  [arXiv:hep-ph/0404168].
  %%CITATION = HEP-PH 0404168;%%

\bibitem{so10inf}
  B.~Kyae and Q.~Shafi,
  %``Inflation with realistic supersymmetric SO(10),''
  Phys.\ Rev.\ D {\bf 72} (2005) 063515
  [arXiv:hep-ph/0504044];
  %  B.~Kyae and Q.~Shafi,
  %``Inflation with a realistic SO(10) model,''
  arXiv:hep-ph/0510300.


\bibitem{dterminf}
  P.~Binetruy and G.~R.~Dvali,
  %``D-term inflation,''
  Phys.\ Lett.\ B {\bf 388} (1996) 241
  [arXiv:hep-ph/9606342].
  %%CITATION = HEP-PH 9606342;%%







\end{thebibliography}
\end{document}